\documentclass[prc,superscriptaddress,unsortedaddress,twocolumn,preprintnumbers,amsmath,amssymb,floatfix,dvipdfmx]{revtex4}
\usepackage{amsmath}
\usepackage{amssymb}
\usepackage{mathrsfs}
\usepackage{bm}
\usepackage{color}
\usepackage{graphicx}
\usepackage{braket}
\usepackage{mathrsfs}
\usepackage{ulem}
\usepackage{here}
\usepackage{url}
\usepackage[
  hidelinks, 
  dvipdfmx   
  ]{hyperref}

\makeatletter
\let\MYcaption\@makecaption
\makeatother
\usepackage{subcaption}
\makeatletter
\let\@makecaption\MYcaption
\makeatother

\makeatletter
\newcommand{\figcaption}[1]{\def\@captype{figure}\caption{#1}}
\newcommand{\tblcaption}[1]{\def\@captype{table}\caption{#1}}
\makeatother

\begin{document}

\title{
Three-body analysis reveals the significant contribution of minor $^{5}$He $s$-wave component in $^{6}$Li$(p,2p)^{5}$He cross section
}

\author{Shoya Ogawa}
\email[]{ogawa.shoya.615@m.kyushu-u.ac.jp}
\affiliation{Department of Physics, Kyushu University, Fukuoka 819-0395, Japan}

\author{Kazuki Yoshida}
\affiliation{Advanced Science Research Center, Japan Atomic Energy Agency, Tokai, Ibaraki 319-1195, Japan}

\author{Yoshiki Chazono}
\affiliation{Department of Physics, Kyushu University, Fukuoka 819-0395, Japan}
\affiliation{RIKEN Nishina Center for Accelerator-Based Science, 2-1 Hirosawa, Wako 351-0198, Japan}

\author{Kazuyuki Ogata}
\affiliation{Department of Physics, Kyushu University, Fukuoka 819-0395, Japan}
\affiliation{Research Center for Nuclear Physics (RCNP), Osaka University, Ibaraki 567-0047, Japan}

\date{\today}

\begin{abstract}
 \noindent
 {\bf Background}:
 $^6$Li is usually treated as an $\alpha+p+n$ three-body system, 
 and the validity of this picture is important for understanding $^6$Li reactions.
 The ($p$,$2p$) reaction is a powerful method to study the structure of valence nucleons in $^6$Li. 
 Recently, the new experimental data of the $^{6}$Li($p$,$2p$)$^{5}$He reaction 
 have been obtained and should be analyzed.
 \\
 {\bf Purpose}:
 We investigate the $^{6}$Li($p$,$2p$)$^{5}$He reaction using the $\alpha+p+n$ three-body wave function 
 of $^6$Li and study the validity of this model.
 \\
 {\bf Methods}:
 We calculate the $^6$Li wave function by using Gaussian expansion method, and the function
 is used to obtain the relative wave function between $p$ and $^5$He.
 We combine the relative wave function with the distorted wave impulse approximation.
 \\
 {\bf Results}:
 Our results reproduce the experimental data of the triple differential cross section
 within about 10\% difference in the absolute values, 
 and contributions from both $p$- and $s$-wave states of $^5$He in $^6$Li are found to be important.
 \\
 {\bf Conclusions}:
 We can qualitatively understand the $^{6}$Li($p$,$2p$)$^{5}$He reaction 
 by describing $^6$Li with the three-body model.
 Contribution from the $s$-wave component is important in reproducing the experimental data
 in the zero recoil-momentum region.
\end{abstract}

\maketitle

\section{Introduction}

$^6$Li is a key nucleus in the big bang nucleosynthesis (BBN), 
and its production amount in the standard BBN model is smaller than the observation result by about three orders, 
which is called the $^6$Li problem. Despite the experimental and theoretical attempts to elucidate 
the production reaction $d+\alpha\to {}^{6}$Li~\cite{Grassi17,Tursunov16,Mukhamedzhanov16,Anders14,Trezzi17,Hammache10}, 
the $^6$Li problem remains unsolved.
$^6$Li has attracted many interests also in nuclear physics, 
mainly because of its simple $d+\alpha$ cluster structure characterized by the small binding energy of about 1.5~MeV. 
One of the constituents, deuteron, is also a weakly-bound nucleus. 
Thus, a deuteron {\lq\lq}cluster'' exists in the $d+\alpha$ cluster structure, 
which causes nontrivial behavior of $d$ inside $^6$Li as pointed out in Refs.~\cite{Watanabe15,Watanabe21}.
Nowadays, not only static properties of $^6$Li, e.g., the binding energy and root-mean-square radius, 
its dynamical properties have also been studied in detail with reaction observables e.g., elastic and breakup cross sections, 
with the $p+n+\alpha$ three-body model~\cite{Kukulin90,Hlophe17,Tursunov16,Christou88,Kikuchi11,Watanabe15,Watanabe21,Bang79,Lehman82,Lehman83,Csoto92,Yamanaka15}.

The ground state of $^6$Li can be characterized as a mixture of the $d+\alpha$ cluster state and the nucleon single-particle (s.p.) state. 
The latter is the main objective of the present study. 
The proton-induced proton knockout reaction, $(p,2p)$, is one of the best tools to pin down the proton s.p. property of nuclei. 
In a recent review~\cite{Wakasa17}, $(p,2p)$ was shown to be an alternative to the electron-induced proton knockout reaction, $(e,e'p)$, 
which is a well-established spectroscopic tool for proton s.p. state~\cite{Kelly96,Kramer01}. 
Recently, new kinematically-complete data of $^6$Li($p,2p$) were obtained at RCNP, Osaka University, with the double arm spectrometer with high resolution~\cite{Yoshida10}.
An important indication of Ref.~\cite{Yoshida10} based on a phenomenological s.p. model of $^6$Li is that the knockout cross section corresponding to the recoilless kinematical condition is sensitive to the proton $s$-wave component in $^6$Li. 
Furthermore, the author controlled the internal energy $\epsilon$ of the $^5$He residue, which is unbound, and discussed the $\epsilon$ dependence of the $s$-wave contribution. 
This idea has also been used in Ref.~\cite{Kubota20} for the study of the neutron s.p. structure of $^{11}$Li as well as the di-neutron correlation, 
in inverse kinematics with more inclusive measurement for the angles of the recoil proton and a knocked-out neutron.
Determination of the $s$-wave probabilities of two-neutron halo nuclei is crucial for understanding their structure.

The purpose of the present work is to perform a distorted wave impulse approximation (DWIA) calculation 
with a sophisticated $^6$Li three-body model used
to clarify whether the predicted $s$-wave probability is consistent with the kinematically complete $(p,2p)$ experimental data.
In Ref.~\cite{Lanen89}, 
a three-body description of $^6$Li was quantitatively justified by the comparison with the $(e,e'p)$ data; 
the overlap function between $^6$Li and $^5$He was calculated in momentum space~\cite{Christou88}. 
As mentioned in Ref.~\cite{Lanen89}, the proton knockout reaction data corresponding to different $\epsilon$ are sensitive 
to the $p$- and $s$-wave mixture in the ground state of $^6$Li. 
Therefore, a stringent test for the three-body model can be done by comparing the DWIA calculation and $(p,2p)$ experimental data.

The construction of this paper is as follows.
In Sec.~II, we describe the theoretical framework. 
In Sec.~III, we present and discuss the numerical results. 
Finally, in Sec.~IV, we give a summary of this study.

\section{Formalism}
\subsection{Gaussian expansion method}

\begin{figure}[tbp]
 \centering
 \includegraphics[scale=0.11]{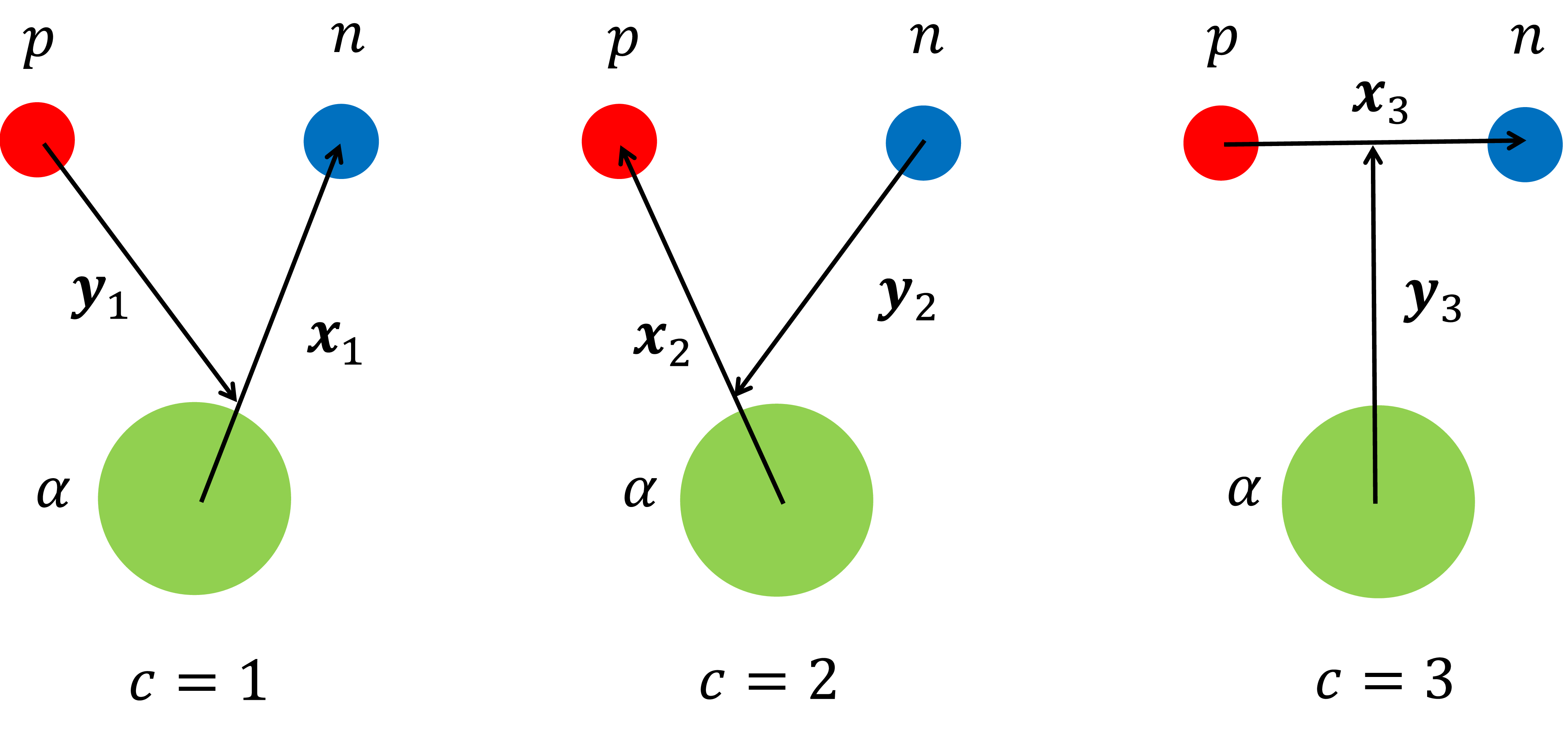}
 \caption{Jacobi coordinate for the three-body system.}
 \label{fig:jacobi}
\end{figure}

We apply the Gaussian expansion method (GEM)~\cite{Hiyama03} to describe the ground state of $^6$Li
and discretized-continuum states of $^5$He.
In GEM,
a wave function of the three-body system is expanded with the Gaussian basis 
on the Jacobi coordinate as shown in Fig.~\ref{fig:jacobi}.
The basis are described as 
\begin{align}
 \psi_{\nu\lambda}(\bm{x}_c) 
  &=
  x^{\lambda}_{c} e^{-(x_{c}/x_{\nu})^2} Y_{\lambda}(\Omega_{x_c}), 
  \\
 \tilde{\psi}_{\mu l}(\bm{y}_c) 
  &=
  y^{l}_{c} e^{-(y_{c}/y_{\mu})^2} Y_{l}(\Omega_{y_c}) 
\end{align}
with
\begin{align}
 x_{\nu} &= (x_{\rm max}/x_{0})^{(\nu-1)/\nu_{\rm max}} ,
  \\
 y_{\mu} &= (y_{\rm max}/y_{0})^{(\mu-1)/\mu_{\rm max}} ,
\end{align}
where the index $\nu$ ($\mu$) means the $\nu$th ($\mu$th) basis function for the Jacobi coordinate
$\bm{x}_{c}$ ($\bm{y}_{c}$), the symbol $\lambda$ ($l$) denotes the angular momentum 
regarding $\bm{x}_{c}$ ($\bm{y}_{c}$). 
Using the basis, we diagonalize the Hamiltonian:
\begin{align}
 \label{eq:hamiltonian_6Li}
 h_{\rm ^{6}Li} &= K_{x} + K_{y} + V_{\alpha p} + V_{\alpha n} + V_{pn} + V_{\rm C} + V_{\rm PF}.
\end{align}
Here, $K_x$ ($K_y$) means the kinetic energy operator associated with $\bm{x}$ ($\bm{y}$).
The interactions for the $p$-$\alpha$,  $n$-$\alpha$, and $p$-$n$ systems are represented as
$V_{\alpha p}$, $V_{\alpha n}$, and $V_{pn}$, respectively,
and $V_{\rm C}$ is the Coulomb interaction between $p$ and $\alpha$.
For the valence nucleons of $^6$Li, the Pauli forbidden state is the 0$s$ orbit occupied by the $\alpha$ core. 
The forbidden state can be excluded from the $\alpha+p+n$ system by using the so-called pseudopotential $V_{\rm PF}$,
which is the same as Eq.~(9) in Ref.~\cite{Watanabe15}.

The Hamiltonian of $^5$He is expressed as
\begin{align}
 \label{eq:hamiltonian_5He}
 h_{\rm ^{5}He} &= K_{x_{1}} + V_{\alpha n} .
\end{align}
Diagonalising this Hamiltonian gives a set of eigenfunctions
$\{\phi_{n\ell_{\rm B} I_{\rm B} M_{\rm B}}\}$,
where $n$, $\ell_{\rm B}$, $I_{\rm B}$, and $M_{\rm B}$ are the radial quantum number, 
the orbital angular momentum, the total spin, and its $z$ component, 
respectively, of the $n$-$\alpha$ system.

We introduce the overlap function defined by
\begin{align}
 \label{eq:overlap}
 \Psi_{n \ell_{\rm B} I_{\rm B}M_{\rm B}M}(\bm{y}_{1}) 
 &\equiv 
 \braket{\phi_{n\ell_{\rm B} I_{\rm B}M_{\rm B}}(\bm{x}_{1})|\Phi_{1M}(\bm{x},\bm{y})}_{\bm{x}_{1}} .
\end{align}
Here, $\Phi_{1M}$ is the ground state wave function of $^6$Li with the total spin $1$ and its $z$ component $M$.
The subscript $\bm{x}_{1}$ of $\braket{\cdots}$ means the integral variable.
Equation~\eqref{eq:overlap} can be expanded as
\begin{align}
 \Psi_{n \ell_{\rm B} I_{\rm B} M_{\rm B}M}(\bm{y}_{1}) 
 &=
 \sum_{\ell j m m_{N}}
 (jm+m_{N} I_{\rm B} M_{\rm B} | 1 M)
 \nonumber \\
 &~~\times
 (\ell m \frac{1}{2} m_{N} | jm+m_{N}) ~
 \varphi_{n\ell j m}(\bm{y}_{1}) \eta_{\frac{1}{2} m_{N}} ,
\end{align}
where $\varphi_{n\ell j m}$ is the bound wave function of the $p$-$^5$He system
with the radial quantum number $n$, the orbital angular momentum $\ell$, 
its $z$ component $m$, and the total spin $j$.
The spin function of the proton is denoted by $\eta_{\frac{1}{2}m_{N}}$,
where $m_{N}$ is the $z$ component of the spin.

\subsection{Quadruple differential cross section}

We apply the DWIA to describe the $^6$Li($p$,$2p$)$^{5}$He reaction.
When a residual $^5$He is in the state $\phi_{\varepsilon\ell_{\rm B} I_{\rm B} M_{\rm B}}$ with
the continuous energy $\varepsilon$, 
the transition matrix is represented as
\begin{align}
 \label{eq:continuous-Tmat}
 T^{\varepsilon\ell_{\rm B} I_{\rm B} M_{\rm B} M}_{m_{1}m_{2}m_{0}}
  =
 \braket{\chi^{(-)}_{1,m_1} \chi^{(-)}_{2,m_2} \phi_{\varepsilon\ell_{\rm B} I_{\rm B} M_{\rm B}}
 |t_{pp}|\chi^{(+)}_{0,m_0} \Phi_{1M}} .
\end{align}
Here, the incident proton is labeled as particle 0, whereas the two emitted protons are as particle 
1 and 2; the antisymmetrization between particles 1 and 2 is taken into account in $t_{pp}$.
A distorted wave of particle $i~(i=0,1,2)$ is described as $\chi_{i,m_{i}}$,
where $m_{i}$ is the $z$ component of the spin of proton.
The outgoing and incoming boundary conditions of the distorted waves are denoted by 
superscripts $(+)$ and $(-)$, respectively.
The $p$-$p$ effective interaction is denoted by $t_{pp}$.
In this study, the spin-orbit part in each optical potential is ignored,
the distorted wave is then expressed as
\begin{align}
 \chi_{i,m_i} = \chi_{i} ~ \eta_{\frac{1}{2}m_{i}} ,
\end{align}
where $\eta_{\frac{1}{2}m_{i}}$ is the spin function of particle $i$.
Inserting the complete set
\begin{align}
  \sum_{n\ell_{\rm B} I_{\rm B} M_{\rm B}}\ket{\phi_{n\ell_{\rm B} I_{\rm B} M_{\rm B}}}
  \bra{\phi_{n\ell_{\rm B} I_{\rm B} M_{\rm B}}} = \hat{1}
\end{align}
into Eq.~\eqref{eq:continuous-Tmat},
the transition matrix can be rewritten as
\begin{align}
 T^{\varepsilon\ell_{\rm B} I_{\rm B} M_{\rm B} M}_{m_{1}m_{2}m_{0}}
  =
  \sum_{n} f^{n}_{\ell_{\rm B} I_{\rm B}}(\varepsilon)
  T^{n\ell_{\rm B} I_{\rm B} M_{\rm B} M}_{m_{0}m_{1}m_{2}}
\end{align}
with
\begin{align}
 f^{n}_{\ell_{\rm B} I_{\rm B}}(\varepsilon)
 =
 \braket{\phi_{\varepsilon\ell_{\rm B} I_{\rm B} M_{\rm B}}|\phi_{n\ell_{\rm B} I_{\rm B} M_{\rm B}}} 
\end{align} 
and
\begin{align}
 \label{eq:dis-Tmat}
 T^{n\ell_{\rm B} I_{\rm B} M_{\rm B} M}_{m_{0}m_{1}m_{2}}
 &=
 \braket{\chi^{(-)}_{1,m_{1}} \chi^{(-)}_{2,m_{2}} |t_{pp}|
 \chi^{(+)}_{0,m_{0}} \Psi_{n\ell_{\rm B} I_{\rm B} M_{\rm B} M}} .
\end{align}

Applying the asymptotic momentum approximation~\cite{Wakasa17,Yoshida16,Yoshida24} to Eq.~\eqref{eq:dis-Tmat},
the quadruple differential cross section (QDX) is expressed as
\begin{align}
 \label{eq:qdx}
 &\frac{d^4 \sigma^{\rm L}}
 {d\varepsilon dE^{\rm L}_{1} d\Omega^{\rm L}_{1} d\Omega^{\rm L}_{2}} 
 \nonumber \\
 &=
 \sum_{\ell_{\rm B} I_{\rm B}} 
 \sum_{\ell j} 
 F_{\rm kin} \mathcal{J}_{\rm LG} 
 \frac{1}{2} \frac{1}{2j+1}
 \left|
 \sum_{n} f^{n}_{\ell_{\rm B} I_{\rm B}}(\varepsilon)
 \bar{T}_{n\ell_{\rm B} I_{\rm B}\ell j}
 \right|^{2} .
\end{align}
Here, $E_{1}$ and $\Omega_{1}$ are the energy and solid angle of particle~1, respectively, 
and $\Omega_{2}$ is the solid angle of particle~2.
The superscript L is attached to the quantities evaluated in the laboratory frame.
The kinematics factor is represented as $F_{\rm kin}$ and 
$\mathcal{J}_{\rm LG}$ is the Jacobian from the center-of-mass frame to the laboratory frame.
The two quantities are the same as those represented in Ref.~\cite{Ogata24}.
It should be noted here that the QDX is an incoherent sum of $\ell_{\rm B}$ and $I_{\rm B}$.
The reduced transition matrix is represented as
\begin{widetext}
 \begin{align}
  \label{eq:dis-Tmat-AMA}
   \bar{T}_{n\ell_{\rm B} I_{\rm B}\ell j}
   =
  \sum_{m_{1}m_{2}m_{0}m_{N}}
   \tilde{t}_{m_{1}m_{2}m_{0}m_{N}}
   \int d\bm{R}~
   \chi^{(-)*}_{1}(\bm{R}) \chi^{(-)*}_{2}(\bm{R})
   \chi_{0}(\bm{R})
   e^{i\bm{K}_{0}\cdot\bm{R}/A}
   \sum_{m} (\ell m \frac{1}{2} m_{N} | j m+m_{N})
   \varphi_{n\ell j m}(\bm{R}) ,
 \end{align}
\end{widetext}
where $\bm{R}$ is the relative coordinate between $^5$He and each particle,
$\bm{K}_{0}$ is the wavenumber of the incident proton,
and $A$ is the mass number of $^6$Li.
The transition matrix of the elementary process is written as
\begin{align}
 \tilde{t}_{m_{1}m_{2}m_{0}m_{N}}
 =
 \braket{\eta_{\frac{1}{2}m_{1}}\eta_{\frac{1}{2}m_{2}} e^{i\bm{\kappa}'\cdot\bm{s}}
 |t_{pp}(\bm{s})|\eta_{\frac{1}{2}m_{0}}\eta_{\frac{1}{2}m_{N}} e^{i\bm{\kappa}\cdot\bm{s}}} .
\end{align}
Here $\bm{s}$ is the relative coordinate between the two protons 
and $\bm{\kappa}$ ($\bm{\kappa}'$) is the relative wavenumber between 
the two protons in the initial (final) state.
Equation~\eqref{eq:dis-Tmat-AMA} is calculated by using the code {\sc pikoe}~\cite{Ogata24}.

We calculate the following triple differential cross section (TDX) for comparison with the experimental data: 
\begin{align}
 \label{eq:tdx}
 \frac{d^3 \sigma^{\rm L}}
 {dE^{\rm L}_{1} d\Omega^{\rm L}_{1} d\Omega^{\rm L}_{2}} 
 =
 \int_{\varepsilon_{\rm min}}^{\varepsilon_{\rm max}} d\varepsilon~
 \frac{d^4 \sigma^{\rm L}}
 {d\varepsilon dE^{\rm L}_{1} d\Omega^{\rm L}_{1} d\Omega^{\rm L}_{2}} .
\end{align}
The interval of integration ($\varepsilon_{\rm min}$, $\varepsilon_{\rm max}$) is determined according to the experimental data.

\section{Results and Discussion}
\subsection{The ground state of $^6$Li}

In this study, we take the Minnesota interaction~\cite{Thompson77} as in Ref.~\cite{Tursunov16}
and the KKNN potential~\cite{Kanada79} for the $p$-$n$ 
and the nucleon-$\alpha$ interactions, respectively, in the calculation of the ground state of $^{6}$Li with GEM. 
To reproduce the binding energy of $^{6}$Li, 
a phenomenological three-body interaction
\begin{align}
 V_{3b}(x,y) = V_{3} e^{-\gamma(x^2+y^2)}
\end{align}
is added to $h_{\rm ^{6}Li}$; we adopt $V_{3}=-0.4$~MeV and $\gamma=0.036$~$\text{fm}^{-2}$.
The parameter sets of the Gaussian basis are summarized in Table~\ref{tab:parameters-GEM}. 
The results of the ground-state energy and root-mean-square radius 
are shown in Table~\ref{tab:energy-rms}, which reproduce well the experimental data~~\cite{Tilley02,Dobrovolsky06}.
\begin{table}[htbp]
 \centering
 \caption{Parameters of Gaussian basis}
 \begin{tabular}{ccccccc} \hline
  c & $i_{\rm max}$ & $x_0$[fm] & $x_{\rm max}$[fm] &
  $j_{\rm max}$ & $y_0$[fm] & $y_{\rm max}$[fm] \\ 
  \hline
  1, 2 & 10 & 0.5 & 12.0 & 10 & 0.5 & 12.0 \\
  3 & 10 & 0.1 & 12.0 & 10 & 0.5 & 12.0 \\
  \hline    
 \end{tabular}
 \label{tab:parameters-GEM}
\end{table}
\begin{table}[htbp]
 \centering
 \caption{
 The ground-state energy and root-mean-square radius of $^6$Li.
 The experimental data are taken from Refs.~\cite{Tilley02,Dobrovolsky06}.
 }
 \begin{tabular}{ccccc} 
  \hline
  & \hspace{5em} Cal. & & \hspace{5em} Exp. & \\
  & $\varepsilon_{0}$ [MeV] & $r_{\rm rms}$ [fm] & $\varepsilon_{0}$ [MeV]
  & $r_{\rm rms}$ [fm] \\
  \hline
  & $-3.77$ & 2.36 & $-3.6989$ & 2.44$\pm$0.07 \\
  \hline    
 \end{tabular}
 \label{tab:energy-rms}
\end{table}

\begin{figure}[tpb]
 \centering
 \includegraphics[scale=0.6]{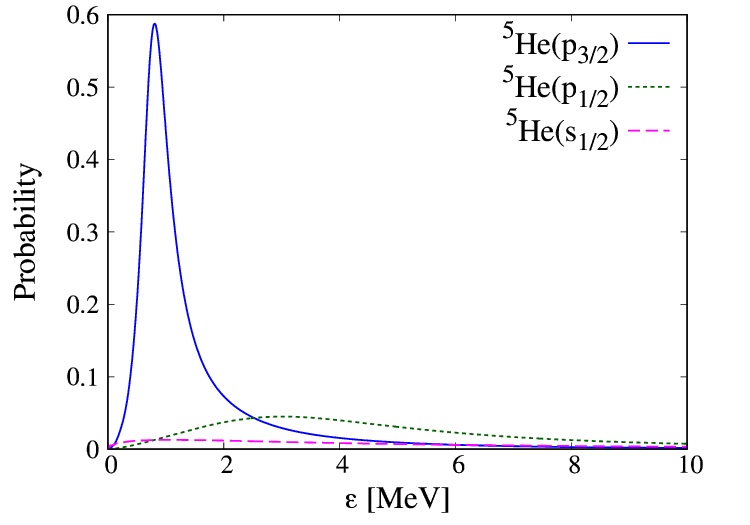}
 \caption{
 Energy distribution of each $^{5}$He state in $^{6}$Li.
 }
 \label{fig:probability-smooth}
\end{figure}

We calculate $^{5}$He energy distribution for individual spin-parity state 
in the ground state of $^{6}$Li:
\begin{align}
 P(\varepsilon)
 =
 \sum_{M_{\rm B}} 
 \int d\bm{y}_{1}~
 \left|
 \sum_{n}f^{n}_{\ell_{\rm B} I_{\rm B}}(\varepsilon) \Psi_{n\ell_{\rm B}I_{\rm B}M_{\rm B}}(\bm{y}_{1})
 \right|^{2} .
\end{align}
We show the value of $P = \int P(\varepsilon) d\varepsilon$ for the $p$- and $s$-wave states of $^5$He in Table~\ref{tab:probability}.
The contributions of higher angular momentum states are found to be negligibly small and not considered in the following discussion.
The $p$-wave states of $^5$He are found to be dominant in $^6$Li.
We also show the energy distribution of the probability in Fig.~\ref{fig:probability-smooth}.
The solid, dotted, and dashed lines represent $P(\varepsilon)$ of $^5$He($p_{3/2}$), 
$^5$He($p_{1/2}$), and $^5$He($s_{1/2}$), respectively.
The probability of $^5$He($p_{3/2}$) and $^5$He($p_{1/2}$) have peak structures around 0.8 MeV and 3 MeV, respectively, 
which are near the resonant energies of the narrow and broad resonances~\cite{Selove74}, respectively.
It indicates that these $^{5}$He resonances exists in our model of $^{6}$Li.
On the other hand, the absolute value of $P(\varepsilon)$ for $^5$He($s_{1/2}$) is very small.
However, as shown in Sec.~III~B, this state makes a significant contribution to the TDX.
\begin{table}[htbp]
 \centering
 \caption{
 $^{5}$He probability of each state in the ground state of $^{6}$Li.
 }
 \begin{tabular}{ccccc}
  \hline
  & ~$^5$He($s_{1/2}$) & ~$^5$He($p_{3/2}$) & ~$^5$He($p_{1/2}$) & \\
  \hline
  $P$ & ~~0.102 & ~~0.551 & ~~0.275 \\
  \hline    
 \end{tabular}
 \label{tab:probability}
\end{table}

\subsection{$^6$Li($p$,$2p$)$^5$He reaction}
\begin{figure}[tpb]
 \centering
 \includegraphics[scale=0.7]{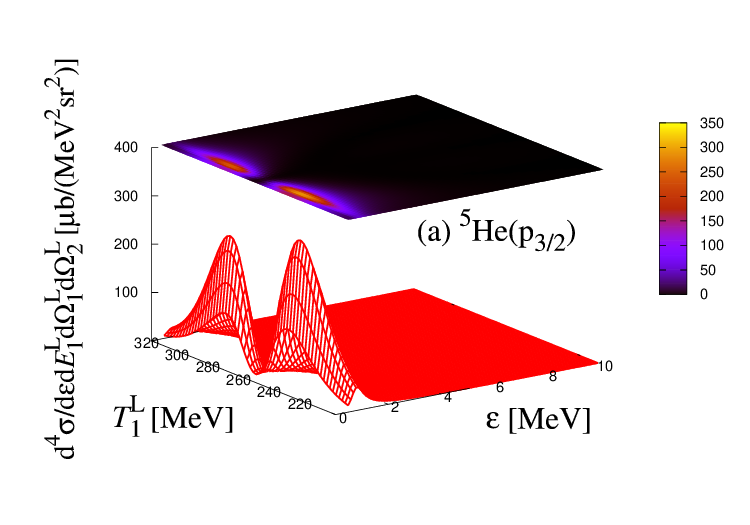}
 \includegraphics[scale=0.7]{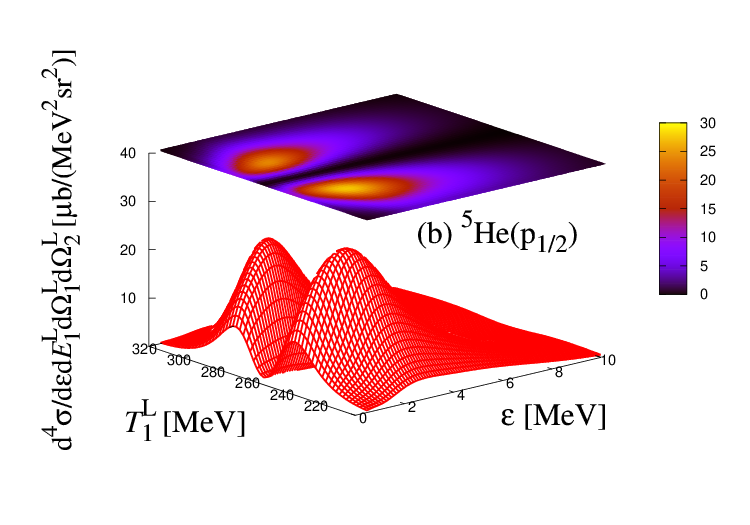}
 \includegraphics[scale=0.7]{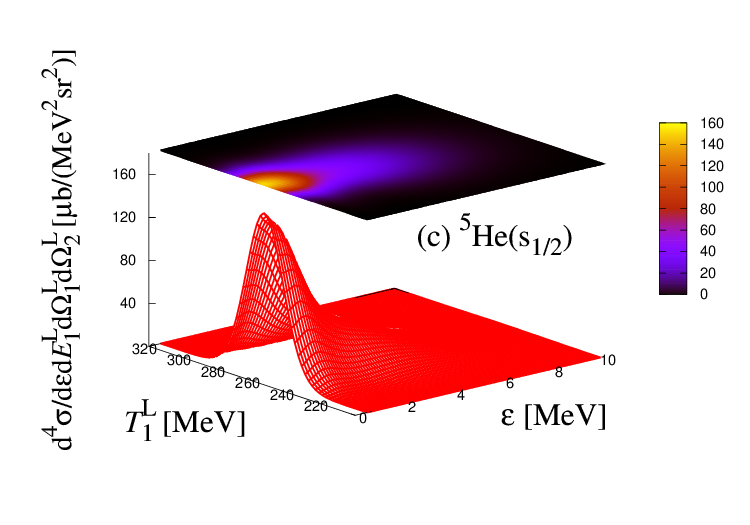}
 \caption{
  QDXs of $^{6}$Li$(p,2p)^{5}$He leaving (a)$^5$He($p_{3/2}$), (b)$^5$He($p_{1/2}$), and
 (c)$^5$He($s_{1/2}$) residue.
 }
 \label{fig:qdx}
\end{figure}
We analyze the $^6$Li($p$,$2p$)$^5$He reaction at 392 MeV.
The polar and azimuthal angles of the particle 1 (2) in the final state are fixed at
$32.21^{\circ}$ and $0^{\circ}$ ($51.62^{\circ}$ and $180^{\circ}$), respectively.
$T^{\rm L}_{1}\simeq260$ MeV satisfies the recoilless condition, 
in which the momentum of $^5$He is almost zero.
We use the Franey-Love effective interaction for $t_{pp}$~\cite{Franey85}.
The optical potentials of the $p$-$^6$Li system is constructed 
by folding the Melbourne g matrix~\cite{Amos00,Minomo10} with the density of $^6$Li. 
The density is obtained by using GEM.
For the optical potentials of the $p$-$^5$He system,
the EDAD1 parameter set of the Dirac phenomenology~\cite{Hama90,Cooper93} is used.
The Coulomb potentials are constructed by assuming a uniformly charged sphere as shown
in Ref.~\cite{Koning03}.
In the present analysis, we investigate the reactions where $p_{3/2}$, $p_{1/2}$, 
and $s_{1/2}$ valence protons are knocked out.
We have confirmed that the contribution from the $d$-wave valence proton is negligibly small.

The QDXs leaving $^5$He($p_{3/2}$), $^5$He($p_{1/2}$), and $^5$He($s_{1/2}$) residues
are shown in Fig.~\ref{fig:qdx} as functions of $\varepsilon$ and 
the kinetic energy $T_{1}^{\rm L}$ of particle 1.
For the $T_{1}^{\rm L}$ distribution, one can see the typical single- and double- peaked shapes of 
$s$- and $p$-wave nucleon knock-outs, respectively.
The QDXs for $^5$He($p_{3/2}$) and $^5$He($p_{1/2}$) have peak structures 
at the resonant energies with respect to $\varepsilon$.
The absolute value of QDX for $^5$He($s_{1/2}$) is larger than that for $^5$He($p_{1/2}$) 
even though $^5$He($s_{1/2}$) has much lower $P$ in $^6$Li.
This can be understood from the difference in the maximum values of the Fourier transform 
of $\varphi_{n\ell jm}$.
In fact, when an $s$-wave proton is knocked out from $^{40}$Ca, 
the TDX tends to have a large absolute value, as seen in Fig.~21 of Ref.~\cite{Wakasa17}.
This is because the momentum distribution of the $s$-wave nucleon is well localized around the zero momentum 
and its peak height can be larger than that for the $p$-waves, 
even though the integrated value of the former is significantly smaller than that of the latter.

\begin{figure}[tbp]
 \centering
 \includegraphics[scale=0.8]{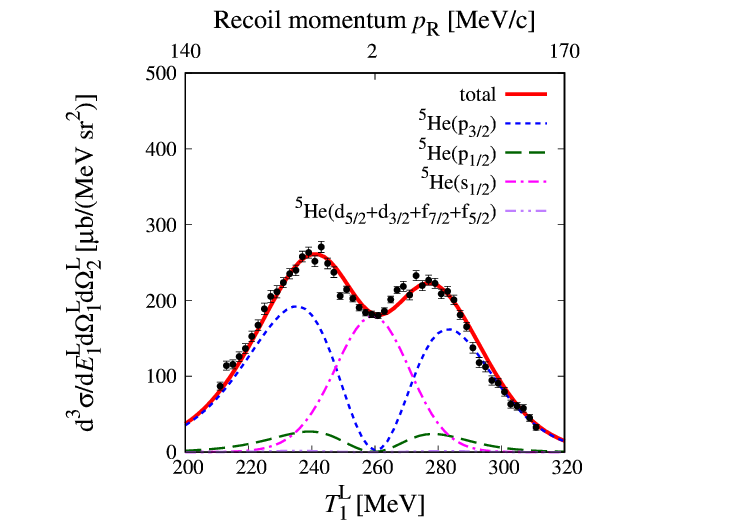}
 \caption{
 TDX calculated by using Eq.~\eqref{eq:tdx} with 
 $({\varepsilon_{\rm min}},{\varepsilon_{\rm max}})=$ (0.4, 1.8).
 The experimental data are taken from Ref.~\cite{Yoshida10}.
 }
 \label{fig:tdx-1}
\end{figure}
\begin{figure}[tbp]
 \centering
 \includegraphics[scale=0.8]{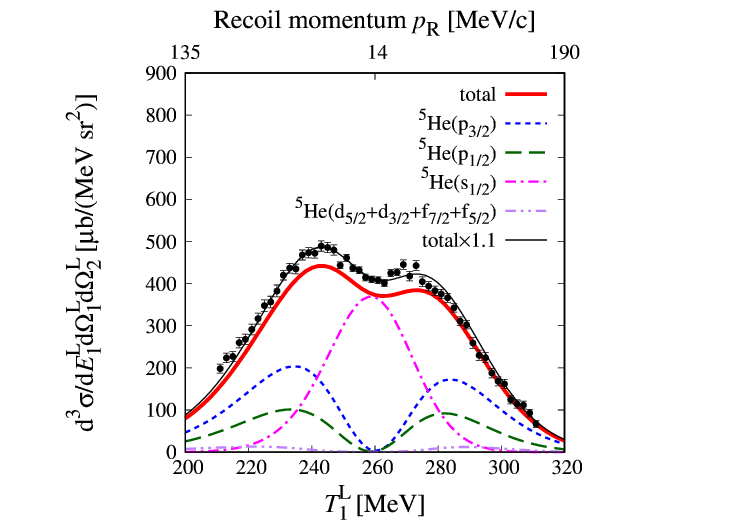}
 \caption{
 Same as in Fig.~\ref{fig:tdx-1} but for
 $({\varepsilon_{\rm min}},{\varepsilon_{\rm max}})=$ (0.4, 14.8).
 }
 \label{fig:tdx-2}
\end{figure}
Finally, we show the TDX calculated by using Eq.~\eqref{eq:tdx} with
$({\varepsilon_{\rm min}},{\varepsilon_{\rm max}})=$ (0.4, 1.8) and (0.4, 14.8)
by the thick lines in Figs.~\ref{fig:tdx-1} and \ref{fig:tdx-2}, respectively.
The dotted, dashed, and dot-dashed lines correspond to the results when the residues are
$^5$He($p_{3/2}$), $^5$He($p_{1/2}$), and $^5$He($s_{1/2}$), respectively.
As mentioned above, the contributions of the $d$- and $f$-waves of $^5$He,
which are represented by the two-dot-dashed lines, are significantly small.
The upper horizontal axes show the recoil momenta $p_{\rm R}$ of $^5$He.
Note that for a given $T^{\rm L}_{1}$, $p_{\rm R}$ depends on $\varepsilon$ 
and the mean values of $p_{\rm R}$ are shown at the dip and the edges of the distribution.
Our results reproduce the shapes of the experimental data~\cite{Yoshida10} well,
but slightly underestimate the absolute value in Fig.~\ref{fig:tdx-2};
we also show the TDX multiplied by a normalization factor by the thin solid line.
Possible origin of this underestimation will be the systematic error of the experimental data, which is estimated about 10\%~\cite{Wakasa23}
because the data has been obtained by using almost the same setup as the data in Refs~\cite{Noro20,Noro23} with 10\% uncertainties.
Another possibility is the uncertainty of the optical potential for the $p$-$^5$He system.
We have also calculated TDXs with $\varepsilon_{\rm min}$ varied by ±10\% around 0.4 MeV to estimate the effect of the uncertainties of $\varepsilon_{\rm min}$ on the TDX,
and confirmed that the TDX changed only a few percent.
In the results for the two energy ranges, the contributions of $^5$He($s_{1/2}$)
to the TDXs at $T_{1}^{\rm L}\sim260$ MeV are significant and fill the dips in the total TDXs. 
The $^{5}$He($s_{1/2}$) contribution becomes more important when we see the TDX corresponding to higher $\varepsilon$. 
The fact that our calculation reproduces well the TDXs for both $\varepsilon$ regions indicates that the proton s.p. nature of $^{6}$Li is properly described with the three-body model adopted. 
It should be noted that in the setup of the $^{6}$Li($p$,$2p$) experiment, 
the finite-size effect of the detectors, which usually affects the TDX around the dip, was minimized [16]. 
We also have confirmed this.

\section{Summary}

We analyzed the $^{6}$Li($p$,$2p$)$^{5}$He reaction by using DWIA and the three-body model.
The $p$-wave states of $^5$He dominate in $^6$Li, 
while the probability of the $s$-wave state is about 10\%.
In TDX analysis, both $p$- and $s$-wave contributions are important,
and in particular, 
the only component contributing to the region around the dip, 
where the recoilless condition is satisfied,
in the experimental data is the $^5$He($s_{1/2}$) component 
while the other components contribution is negligible.
Our results reproduce the experimental data of the TDX
within about 10\% difference in the absolute values.
These results indicate that we can qualitatively describe the s.p. property of $^{6}$Li, which is consistent with 
the $^{6}$Li($p$,$2p$)$^{5}$He reaction data by using an $\alpha+p+n$ three-body model.
As a next step, the $\alpha+d$ structure of $^{6}$Li should be judged. 
For this purpose,
we will analyze the $^{6}$Li($p$,$pd$)$\alpha$ reaction by combining the three-body model with CDCCIA~\cite{Chazono22},
which is a new reaction model to describe the reaction for knocking out a weakly bound cluster.

\section*{Acknowledgments} 

The authors thank H.~P.~Yoshida and T.~Noro for providing them with the experimental data.
S.O. and K.O. thank T.~Wakasa for helpful discussions.
This work is supported in part by Grant-in-Aid for Scientific Research
(No.\ JP20K14475, and No.\ JP21H04975)
from Japan Society for the Promotion of Science (JSPS) and JST ERATO Grant No. JPMJER2304, Japan.

\bibliography{./ref}

\end{document}